# Swiss Elections to the National Council:

# First trials with e-voting in elections at federal level

**By Anina Weber (anina.weber@bk.admin.ch) and Geo Taglioni (geo.taglioni@bk.admin.ch), Swiss Federal Chancellery**

**Dagstuhl-Seminar 11281: Verifiable Elections and the Public (10.-15.07.2011)**

**Organizers:** Michael Alvarez (CalTech - Pasadena, US), Josh Benaloh (Microsoft Corp. - Redmond, US), Alon Rosen (The Interdisciplinary Center - Herzliya, IL), Peter Y. A. Ryan (University of Luxembourg, LU)

On October 23$^{rd}$ 2011, around 22'000 voters will be authorized to cast their votes electronically in occasion of the elections to the National Council. These are the **first trials ever with e-voting in elections at federal level in Switzerland**. Four cantons are going to conduct trials with this new channel. Only Swiss voters living abroad will be authorized to participate.

The Swiss Confederation pursues the long term goal of the introduction of e-voting as a **third, complementary voting method** – in addition to voting in person at the polling station and postal voting.

1. Project History

The Swiss people **trust** the authorities that polls are conducted legally. This trust made it possible to introduce **postal voting** in 1994. Since then Swiss voters automatically receive postal vote documents and may cast their vote by post without requiring any special reason to do so. Nowadays, this more convenient way to cast votes remotely is established: more than 90% of the votes are cast by post. The new channel was able to stop the sinking participation rate, more so it had a slightly positive effect on voter participation. The postal vote has been the **starting point** for the development and the introduction of electronic voting.

**E-voting** is a further development of opportunities for participation. In Switzerland, it is the **third voting channel**, an instrument for the society of the 21st Century, where new technologies, but also the mobility of people are becoming increasingly important. Since 2000, e-voting has been started in **three pilot cantons** (Geneva, Neuchatel, and Zurich). The **Confederation** contributed with funds for the development of the three systems, one per pilot canton, and exercised a **coordination role**. The three systems take into account the differences between the political systems of the cantons.

After an initial testing phase, the Federal Council and Parliament have decided of 2006/07 a **gradual and controlled expansion of e-voting**. In this testing-phase, the **Swiss voters abroad** have been identified as the group with the greatest benefits. E-voting meets the needs of the approximately 130'000 registered Swiss voters abroad very well, since they often face problems with the postal services. They also form a relatively small group, which suits well for a testing phase. For these reasons, Swiss abroad are treated with priority.

Since 2010, further cantons started with e-voting trials. The expansion to include **13 cantons** has been possible thanks to the idea of the **"system hosting"**. Cantons wanting to introduce e-voting can use the existing systems developed by one of the pilot canton. This is a convenient and economically attractive way for them to gain experiences with the new voting channel: the new cantons can benefit from the existing expertise made by the pilot cantons.

Since 2004, almost **twenty trials** have been successfully conducted at **federal level** and many more on cantonal and communal level. In the context of the most recent trial in February 2011, approximately 177'500 voters were authorised to participate. Some 25'600 voters cast their votes electronically, which corresponds a participation of 14.4 %.

On October 23rd 2011, the **first ever trials with e-voting in elections** at federal level[1] will be performed. Four cantons (Basel, Aargau, St.Gallen, and Grisons) will participate at these trials. Around 22'000 Swiss voters living abroad will be authorized to cast their votes electronically. Two systems are being used for the elections: the systems of Geneva and Zurich.

## 2. The Legal Principles of E-Voting

The legal principles of e-voting in Switzerland are found in the Federal Constitution, the Federal Act on Political Rights, and the Federal Ordinance on Political Rights.

### a) Federal Constitution (art. 34)

Article 34 of the Federal Constitution guarantees the political rights. This guarantee protects the freedom of the citizen to form an opinion and to give genuine expression to his or her will.

### b) Federal Act on Political Rights (art. 8*a*)

The Federal Council may in consultation with interested cantons and communes permit electronic voting pilot schemes that are limited in their geographical scope, in the dates on which they are held, and in the subject matter to which they relate. It may make authorisation subject to requirements or conditions or, taking account of the overall circumstances, exclude electronic voting at any time, whether in terms of its geographical scope, the subject matter to which it relates, or the date on which it is held.

---

[1] Elections to the National Council.

The verification of eligibility to vote, voting secrecy and the counting of all the votes cast must be guaranteed and abuses prevented.

**c) Federal Ordinance on Political Rights (art. 27*a* ff. PoRO)**

In the Federal Ordinance on Political Rights, 19 articles are dedicated to e-voting. The most important provisions are the following:

- **Art. 27*a*:** Federal Council authorisation is required for the conduct of e-voting trials at federal level.

- **Art. 27*c* para. 2:** Limitation to 10% of the entire Swiss electorate and to 20% of the cantonal electorate (not including Swiss voters living abroad)

- **Art. 27*d*** (Requirements):

  – Only persons who are eligible to vote may cast a vote (verification of eligibility);

  – Each person who is eligible to vote has only one vote and may vote only once (uniqueness of the vote);

  – Third parties must not be able, systematically and effectively, to intercept, alter or divert votes cast electronically (reliable expression of the voter's genuine intention);

  – Third parties must be unable to find out how a person has voted (voting secrecy);

  – All votes must be counted to determine the result (reliability of the count);

  – It must be possible to prevent systematic abuse (compliance with voting procedures).

### 3. Federalism and the three e-voting systems

The Swiss Confederation regulates the exercise of political rights in federal matters, the Cantons regulate their exercise in cantonal and communal matters (Art. 39 para. 1 Constitution). The Confederation specifies when a vote will be held and what proposals are voted on, while the cantons are responsible for the organisation and conduct of the vote (e.g. Art. 10 PoRA). The Confederation lays down only the ground rules that must be applied in all the cantons, while the cantons regulate the details of procedures for exercising political rights in accordance with their own rules on voting.

Due to the Swiss **federalism** and the **different cantonal political systems**, three different e-voting systems have been developed in Switzerland:

- The **Geneva system** offers a centralized solution. The canton is in the foreground. Geneva is the owner of the software and the operator of the system as well.

- The **Zurich system** is based on a decentralized solution. The communes are in the foreground. The canton is the owner of the software. An external company operates the system.

- The **Neuchâtel system**, as the Geneva system, operates on a centralized solution. The canton is the owner of the software and operates the system originally developed by a private firm. The e-voting system of Neuchâtel is integral part of the so-called "Guichet Unique". This online counter offers the opportunity to perform many other legal acts – such as the tax declaration – via the Internet.

All systems are conceived to operate in the **four Swiss national languages** (German, French, Italian, and Romanic).

### 4. The Role of the Federal Chancellery

Actual project managers for the introduction of e-voting are the **cantons**. They are free to decide whether and when they want to introduce e-voting and they are responsible for the (technical) arrangements. The Federal Chancellery has the overall responsibility for the project at the federal level. It prepares the **authorisation procedure** by the Federal Council and **coordinates** the cantonal projects. The Chancellery devises and develops the **legal principles** and defines **best practices** for e-voting as well. It advises the cantons and has the lead in the **communication** at the federal level.

### 5. Current work

**a) Political/organisational matters**

A **steering committee** as well as a **support group** for e-voting have recently been set up. The aim is a broader support for the project within the Confederation and the cantons as well as an improved cooperation on strategic matters.

The Federal Chancellery pleads for an ongoing **expansion** of e-voting to include new cantons and supports those wishing to introduce the new voting channel.

In 2012/13, a **third e-voting report** will be drafted. This report aims to evaluate the experiences since the last report of 2006 and to identify the next steps as well as the necessary legal adoptions.

Switzerland also intends to continue its involvement at the **international level** in the field of e-voting.

In summer 2011, a joint federal/cantonal **communication and media concept** has been elaborated. The goal is increased information for citizens, politicians, media and academia. One of the measures identified is a **brochure** on the project which is currently being distributed.

**b) Security matters**

A **working group** devising **minimum technical security standards** for e-voting (e.g. question of certification) has been set up in spring 2011.

The Federal Chancellery also collaborates with the **research**: the ETH Zurich currently works on the question of secure voting by an insecure client (results expected by 2013) and the Bern University of Applied Sciences recently conducted a study on the feasibility and the costs of verifiable e-voting systems.

## 6. Future Goals

The Federal Chancellery seeks the following goals:

- In the **medium term**, the vast majority of Swiss abroad should be able to vote electronically in federal popular votes by 2012 and in federal elections by 2015.

- In the **long term**, e-voting should be introduced as a third, complementary voting method in addition to voting in person at the polling station and postal voting. Everyone who is eligible to vote should be able to cast their vote per internet.

More information about the Swiss e-voting project can be found on www.bk.admin.ch → "Themen" → "Vote électronique" (German, French, Italian, and Romanic).